\documentclass[oneside,twocolumn,aps,prl,preprintnumbers,bibnotes10pt,superscriptaddress,amsmath,amssymb]{revtex4-2}

\usepackage{graphicx}
\usepackage{subcaption}
\usepackage{float}
\usepackage[dvipsnames]{xcolor}
\usepackage{epstopdf}
\usepackage{amsmath}
\usepackage{braket}
\usepackage{amsthm}
\usepackage{amssymb}
\usepackage{amsfonts}
\usepackage{txfonts}
\usepackage{mathtools}
\usepackage{bm}
\usepackage{hyperref}
\usepackage{lipsum}
\usepackage[sort&compress]{natbib}
\usepackage{comment}
\usepackage{physics}
\usepackage{cancel}

\usepackage{lipsum}
\usepackage{booktabs}
\usepackage[normalem]{ulem}

\usepackage{tikz}
\usetikzlibrary{quantikz2,calc,arrows,chains,matrix,positioning,scopes}

\usepackage{letltxmacro}
\LetLtxMacro{\originaleqref}{\eqref}
\renewcommand{\eqref}{Eq.~\originaleqref}

\newcommand{\probP}{\text{I\kern-0.15em P}}

\usepackage{tikz}
\makeatletter
\newcommand\mathcircled[1]{%
  \mathpalette\@mathcircled{#1}%
}
\newcommand\@mathcircled[2]{%
  \tikz[baseline=(math.base)] \node[draw,circle,inner sep=0.5pt] (math) {$\m@th#1#2$};%
}
\makeatother

\begin{document}

\title{Dynamical universality class for competing short- and long-range interactions}

\author{Jean-Fran\ifmmode \mbox{\c{c}}\else \c{c}\fi{}ois de Kemmeter}
\affiliation{Istituto Nazionale di Ottica (INO), Consiglio Nazionale delle Ricerche (CNR), Largo Enrico Fermi 6, I-50125 Firenze, Italy.}
\affiliation{Department of Mathematics, Royal Military Academy, Brussels, Belgium.}

\author{Stefano Ruffo}
\affiliation{Istituto dei Sistemi Complessi (ISC), Consiglio Nazionale delle Ricerche (CNR), Via Madonna del Piano 10, I-50019 Sesto Fiorentino, Italy.}
\affiliation{INFN, Sezione di Firenze, I-50019 Sesto Fiorentino, Italy.}
\affiliation{SISSA, via Bonomea 265 I-34136 Trieste, Italy.}

\author{Stefano Gherardini}
\affiliation{Istituto Nazionale di Ottica (INO), Consiglio Nazionale delle Ricerche (CNR), Largo Enrico Fermi 6, I-50125 Firenze, Italy.}

\begin{abstract}
Understanding the dynamical universality classes of systems with long-range interactions remains a key challenge in statistical physics. In this Letter, we analytically and numerically investigate the non-equilibrium critical dynamics of the one-dimensional spin-$1/2$ Nagle-Kardar model, which is characterized by the competition between short- and long-range interactions and the presence of a tricritical point. We focus on the slowing-down of the magnetization $m$ at criticality under Glauber dynamics. Starting from the corresponding master equation, we perform a coarse-graining procedure to obtain a Fokker-Planck equation for the macroscopic variables. Then, the asymptotic decay of the magnetization is derived using central manifold theory. We find that $m$ decays as $t^{-1/2}$ along the critical line and as $t^{-1/4}$ precisely at the tricritical point. This finding confirms that the dynamical critical exponent is $z=2$ as for mean-field models, proving that the macroscopic critical dynamics of the Nagle-Kardar model falls within the dynamical universality class of purely relaxational, non-conserved order parameters (model A). While Kardar proved that the equilibrium Curie-Weiss theory extends to Ising models where nearest-neighbor interactions are included, we here show that such result is valid also for critical dynamics. Our work provides the semi-analytical solution for the critical dynamics of a model with mixed-range interactions, assigning its universality class.
\end{abstract}

\maketitle

Long-range classical and quantum interacting systems~\cite{barre2001inequivalence,campa2009statistical,campa2014physics,defenu2023long} are characterized by an inter-particle potential $V(r)$ that decays as $V(r) \sim r^{-\alpha}$ at large separation $r$. Examples include Coulomb interactions~\cite{levin2014nonequilibrium}, self-gravitating systems~\cite{chavanis2006phase}, two-dimensional turbulence~\cite{bouchet2012statistical}, and Riesz gases~\cite{lewin2022coulomb,krapivsky2025expansion}.

In many real-world systems short- and long-range interactions coexist and can compete. The addition of a mean-field term to the Ising Hamiltonian was first proposed by Baker~\cite{baker1963ising} and later analyzed in the canonical ensemble by Nagle~\cite{nagle1970ising} and Kardar~\cite{kardar1983crossover}. In the phase diagram of this model there are lines of first and second-order phase transitions separated by a tricritical point~\cite{mukamel2005breaking}. The Nagle-Kardar model can represent such diverse physical systems as: bond percolation in the presence of infinite-range bonds~\cite{kaufman1984short}, spin systems on small-world hypergraphs~\cite{bolle2006thermodynamics}, single-chain magnets~\cite{gatteschi2014single}, synchronization with long-range interactions~\cite{gupta2017spontaneous}, and bosons in an optical lattice with cavity-induced long-range interactions~\cite{blass2018quantum}. Recently, the Nagle-Kardar model has been used to characterize finite-size effects, like the Casimir force, near criticality~\cite{dantchev2024fluctuation,Dantchev2024,Dantchev2025Casimir}, and the addition of next-to-nearest neighbor and biquadratic interactions has been also considered~\cite{yang2022effect,campa2019ising,campa2025ensemble}.

While the static properties of the Nagle-Kardar model have been widely studied, dynamical properties have not been addressed carefully, if we exclude a single paper where simulations of the dynamics in the microcanonical ensemble were performed~\cite{mukamel2005breaking}. On the other hand, critical slowing down has been investigated for the Ising~\cite{hohenberg1977theory,SchneiderPRL1972,wang1995study} and the Curie-Weiss~\cite{griffiths1966relaxation,ColonnaRomanoPRE2014,aiudi2023critical} models separately.

In this Letter, we study the dynamics of the one-dimensional Nagle-Kardar model in the canonical ensemble, using numerical simulations and analytical tools. Our main contributions are: \textit{i)} a Fokker-Planck equation for macroscopic variables from the microscopic Glauber dynamics~\cite{kikuchi1991metropolis}, for a detailed derivation see the companion paper~\cite{our_paper_long}; \textit{ii)} the confirmation of the dynamical critical exponent $z=2$ that governs the algebraic slowing-down of the magnetization along the second-order critical line and at the tricritical point. This proves that the addition of nearest-neighbor interactions to the mean-field term does not modify the dynamical universality class. Rather than relying on the standard Landau-Ginzburg framework to determine the dynamical exponent, we derive it from first principles by systematically coarse-graining the underlying microscopic dynamics.

Our starting point is the exact master equation describing the stochastic temporal evolution of spin configurations driven by Glauber dynamics~\cite{binder1992monte,landau2021guide}. In order to reduce the system effective dimensionality, we perform a coarse-graining procedure by classifying the configurations in terms of two macroscopic variables: the total magnetization $M$ and the number of defects $2S$ that separate adjacent, oppositely aligned spins~\footnote{The fact that the number of defects is even is the consequence of choosing periodic boundary conditions.}. Then, we derive the master equation that describes the time evolution of the joint probability distribution $P_N(M,2S;t)$, where $N$ is the system size. In the large-$N$ limit, we obtain the Fokker-Planck equation for the time evolution of the joint probability distribution $p(m,s;t)$ in terms of the intensive variables $m = M/N \in[-1,1]$ and $s = 2S/N$. From the Fokker-Planck equation, we determine the asymptotic dynamics of the magnetization $m$. Based on considerations from scaling theory~\cite{FisherPRL1972,Barber1983,Privman1990Editor,wang1995study,ma2018modern}, the average absolute magnetization is expected to decay algebraically at criticality, both in time and with respect to $N$:
\begin{equation}\label{eq:scaling_relations}
    \left\langle \left\vert m(t) \right\vert \right\rangle \sim t^{-\lambda_m} \quad \text{and} \quad \left\langle \left\vert m \right\vert \right\rangle \sim N^{-\Delta_m},
\end{equation}
where the latter holds at equilibrium, $t \to \infty$. We derive that $\lambda_m = 1/2$ along the critical line, and $\lambda_m = 1/4$ at the tricritical point. The exponent $\lambda_m$ is related to the static exponents $\beta,\nu$ and to the dynamical exponent $z$ by the relation $\lambda_m = \beta/(z\nu)$~\cite{wang1995study}. We show that this implies $z=2$, which is the dynamical critical exponent of the Model A dynamic universality class~\cite{hohenberg1977theory}, which describes Markovian critical dynamical phenomena with a non-conserved order parameter. Moreover, $\Delta_m=1/4$ for the critical line~\cite{Ellis2010Asymptotic,ColonnaRomanoPRE2014}, and $\Delta_m = 1/6$ at the tricritical point~\cite{our_paper_long}. This result follows from
\begin{equation}\label{eq:new_scaling_law}
\Delta_m = \frac{ \beta }{ \nu \, d_{u} },
\end{equation}
where $d_{u}$ is the upper critical dimension. \eqref{eq:new_scaling_law} also implies that the Nagle-Kadar model falls into the dynamical universality class of \emph{mean-field} models, since the critical exponent $\Delta_m$ is obtained using the relation $\Delta_m=\beta/(\nu d)$ and replacing $d$ with the upper critical dimension (see Appendix A). For the Curie-Weiss model, \eqref{eq:new_scaling_law} has been verified in Refs.~\cite{Ellis2010Asymptotic,ColonnaRomanoPRE2014} with $d_u = 4$. Here, we extend this result to the Nagle-Kardar model where $d_{u}=4$ on the critical line, and $d_{u}=3$ at the tricritical point~\cite{WegnerPRB1972}. We further discuss the validity of \eqref{eq:new_scaling_law} in the Joint Paper  Ref.~\cite{our_paper_long}. 

\begin{figure}[t!]
    \centering
    \includegraphics[width=0.95\linewidth]{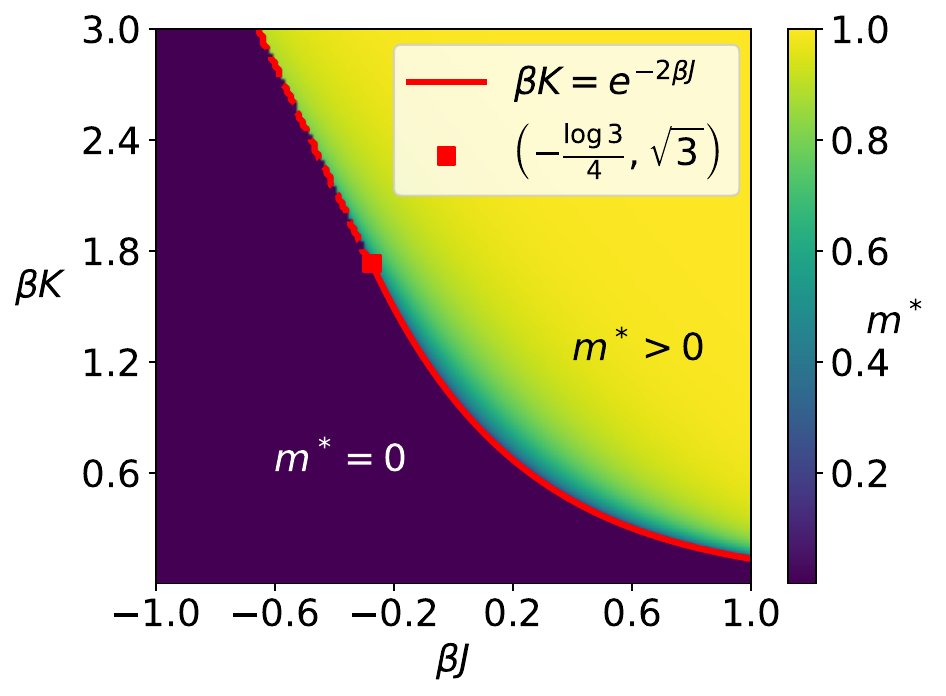}
    \caption{
    Phase diagram of the Nagle-Kardar model \eqref{eq:Hamiltonian} in the canonical ensemble for $h=0$. The red line separates ferromagnetic from paramagnetic states. A tricritical point (red square) marks the junction between a continuous (second-order) transition line (solid red) and a discontinuous (first-order) one (dashed red).
    }
    \label{fig:Phase_diagram}
\end{figure}

\textit{Equilibrium properties of the Nagle-Kardar model in the canonical ensemble.}---We consider a one-dimensional periodic chain of $N$ spins $s_i\in \{\pm 1\}$, described by the Hamiltonian
\begin{equation}\label{eq:Hamiltonian}
    H = -J \sum_{i=1}^N s_i s_{i+1} - \frac{K}{2N} \left(\sum_{i=1}^N s_i\right)^2 - h \sum_{i=1}^N s_{i}\,,
\end{equation}
where $J, K$ are the nearest-neighbor and mean-field coupling respectively, and $h$ denotes the external field. Periodic boundary conditions imply $s_{N+1} = s_1$. The total magnetization of the chain is defined as $M =\sum_{i=1}^N s_i$, while the (even) number of defects is $2S=\sum_{i=1}^N (1-s_i s_{i+1})/2$. Using $M$ and $S$, the Hamiltonian (\ref{eq:Hamiltonian}) can be rewritten as 
\begin{equation}\label{eq:Hamiltonian_v2}
H = -J (N-4S) - \frac{K}{2N} M^2 - h M.
\end{equation}
As $N\rightarrow +\infty$, the equilibrium values of the magnetization per site $m^*$ and the fraction of defects $s^{*}$ satisfy the following implicit equations~\cite{kardar1983crossover}:
\begin{equation}
\begin{split}
    m^{*} &=\frac{\sinh\Big( (h+Km^*)/T \Big)}{\sqrt{\sinh^2\Big( (h+Km^*)/T \Big) + e^{-4 J/T}} },\\
    s^* &= \frac{ 1 - \sqrt{{m^*}^2 + e^{4 J/T} (1 - {m^*}^2)}}{1-e^{4 J/T}},
    \label{eq:ms}
    \end{split}
\end{equation}
where $T$ is temperature and the Boltzmann constant $k_B$ is set to $1$. As shown in Fig.~\ref{fig:Phase_diagram}, the phase diagram of the Nagle-Kardar model in the plane $h=0$ exhibits a tricritical point at $J/T=-\log(3)/4$ and $K/T =\sqrt{3}$~\cite{nagle1970ising, kardar1983crossover}. At the tricritical point, a line of second-order phase transition meets a line of first-order transition.

Besides equilibrium properties for averages, results for the fluctuations of the macroscopic quantities can be obtained using large-deviation techniques~\cite{our_paper_long}. For instance, in the paramagnetic phase and away from the second-order transition line $K/T = e^{-2 J/T}$, $\vert m^* \vert \sim N^{-1/2}$. At the second-order transition line, $\vert m^* \vert \sim N^{-1/4}$, while at the tricritical point $\vert m^* \vert \sim N^{-1/6}$.

\textit{Dynamics: Master equation for the Glauber dynamics.}---We explore the time evolution of spin configurations using the Glauber dynamics~\cite{glauber1963time}. Starting from a configuration $\bm{\sigma}$, a spin is selected uniformly at random and flipped with acceptance probability $1/(1+e^{\Delta E/T})$, where $\Delta E$ is the energy change associated with the flip of the selected spin. In the limit of continuous time, the probability $\probP(\bm{\sigma};t)$ to observe a configuration $\bm{\sigma}$ at time $t$ is ruled by the master equation
\begin{equation}\label{eq:Glauber}
    \frac{d \probP(\bm{\sigma};t)}{dt} = \sum_{\bm{\sigma'}}  \Big[
    \mathrm{T}_{\bm{\sigma}\bm{\sigma'}} \probP(\bm{\sigma'};t)
    -\mathrm{T}_{\bm{\sigma'}\bm{\sigma}} \probP(\bm{\sigma};t)
    \Big],
\end{equation}
where $\mathrm{T}_{\bm{\sigma}\bm{\sigma'}}$ is the Markovian transition rate for passing from the configuration $\bm{\sigma'}$ to the configuration $\bm{\sigma}$. These rates form the entries of a $2^N \times 2^N$ transition matrix of which most of the entries evaluate to zero. 

\begin{figure}[t]
    \centering
    \includegraphics[width=0.95\linewidth]{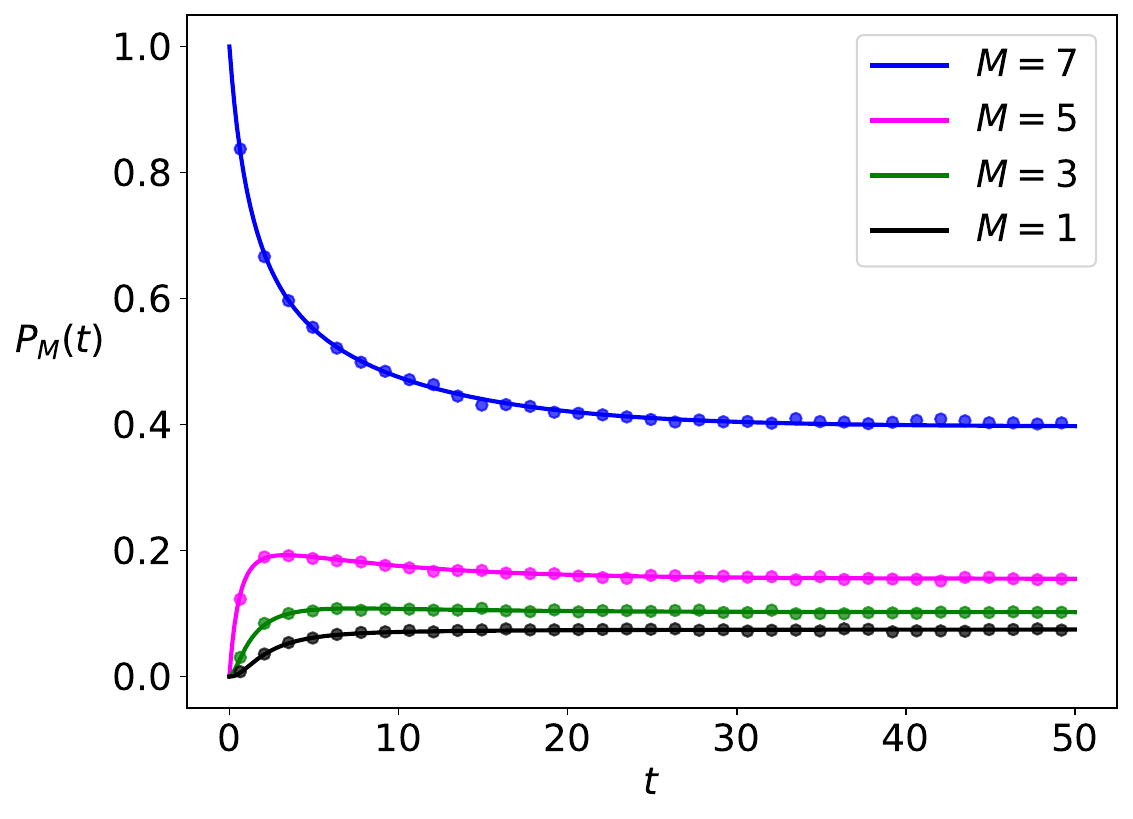}
    \caption{ 
    Comparison between the Glauber dynamics of \eqref{eq:Glauber} (points) and the coarse-grained dynamics  of \eqref{eq:dpdt} (solid lines), with parameters $T=1, h=0.1, J=0.5, K=0.4$ and with $N=7$ spins. We plot the probability $P_M(t)$ that $M$ takes a given value, as a function of time $t$, given by the Monte Carlo sweeps~\cite{bookBinder}. In order to compute the probabilities of the Glauber dynamics we calculate the histogram of $M$ for every time step, using $100$ samples. For the coarse-grained dynamics, the probability $P_M(t)$ is obtained by solving the system of ordinary differential equations of \eqref{eq:dpdt}. For the purpose of the presentation, only positive values of $M$ are plotted. 
    }
    \label{figEM:Glauber_vs_coarsegraining}
\end{figure}

In order to reduce the complexity of the dynamics, we adopt a coarse-grained description of the configurations, whose dynamics is governed by the joint probability distribution $P_N(M,2S;t)$ in terms of $M$ and $2S$. To derive the time evolution of $P_N(M,2S;t)$, we analyze how single spin flips affect the macroscopic variables $M$ and $2S$. Given a spin configuration $\bm{\sigma}$ with magnetization $M$ and $2S$ defects, flipping a single spin results in a new configuration $\bm{\sigma'}$ with updated magnetization $M'\in \{M\pm 2\}$ and number of defects $2S'\in\{2S,2S\pm 2\}$. The corresponding dynamics is described by the master equation
\begin{eqnarray}\label{eq:dpdt}
&&\frac{d P_N(M,2S;t)}{d t} \nonumber \\
&& = - \sum_{\substack{\Delta M =  \pm 2 \\ \Delta S = 0, \pm 2}} \mathcal{T}(M\!+\!\Delta M,2S \!+\! \Delta S \vert M,2S) P_N(M,2S;t) \nonumber\\
&& + \!\!\!\! \sum_{\substack{\Delta M =  \pm 2 \\ \Delta S = 0, \pm 2}} \!\!\!\! \mathcal{T}(M,2S\vert M\!+\!\Delta M,2S \!+\! \Delta S)  P_N(M\!+\!\Delta M,2S \!+\! \Delta S;t).\quad
\end{eqnarray}
The transition rates $\mathcal{T}$ are products of two factors: \textit{(i)} the probability of choosing a specific spin within a configuration, and \textit{(ii)} the Glauber acceptance probabilities. While the latter factor depends only on the energy change and hence only on the change in $M$ and $S$, the probability of choosing a spin depends, a priori, on the specific features of configurations. On the other hand, when deriving the coarse-grained transition rates leading to \eqref{eq:dpdt}, we \textit{assume} that the system dynamics explores the microstates sufficiently fast compared to the relaxation time of the macroscopic variables $(M,S)$ appearing in Hamiltonian (\ref{eq:Hamiltonian_v2}). The validity of this coarse-graining procedure can be appreciated in Fig.~\ref{figEM:Glauber_vs_coarsegraining}, where we compare the time evolution of the probability distribution obtained from the Glauber dynamics (dots) with the results from the coarse-grained dynamics of $P_N(M,2S;t)$ (solid lines). Even for small system sizes $(N=7)$, the agreement is highly satisfactory. The \textit{assumption} above is consistent with the smoothness of the time-evolution of $P_M(t)$.

We now describe the form of the transition rates $\mathcal{T}$. Taking the specific rate
$\mathcal{T}(M-2,2S \vert M,2S)$ as an example, we explain the meaning of each of its terms. This rate corresponds to the transition that flips a spin $+1$ initially surrounded by a single defect, which makes the magnetization decrease by $2$ while keeping the number of defects unchanged. Hence, it can be decomposed as follows:
\begin{equation}\label{eq:single_transition_rate}
    \mathcal{T}(M-2,2S \vert M,2S) = A_{M\rightarrow M-2}^{2S\rightarrow 2S}
    B_1^{+}(N,M,2S),
\end{equation}
where $A_{M\rightarrow M-2}^{2S\rightarrow 2S}$ is the Glauber acceptance probability of a spin flip that changes the magnetization from $M$ to $M-2$ leaving unchanged the number of defects, while $B_1^{+}(N,M,2S)$ is the probability that the selected spin is $+1$ (identified by the superscript $+$) surrounded by a single defect (subscript $1$). The Glauber acceptance probability can be readily obtained by computing the energy change associated to $M\rightarrow M-2$ and $2S \rightarrow 2S$ of the macroscopic variables, using the Hamiltonian (\ref{eq:Hamiltonian_v2}). It reads as
\begin{equation}\label{eq:Glauber1}
    A_{M\rightarrow M-2}^{2S\rightarrow 2S} = \frac{
    e^{- \frac{1}{T}\left[ \frac{K(M-1)}{N} + h \right]}
    }{
    2\cosh \left\{ \frac{1}{T}\left[ \frac{K(M-1)}{N} + h \right] \right\}
    }.    
\end{equation}
For periodic boundary conditions, the coefficient $B_1^{+}(N,M,2S)$ is given by
\begin{equation}\label{eq:B1_plus}
    B_1^{+}(N,M,2S) = \sum_{k\geq 0} k \, \frac{\Omega_1^{+}(N,M,2S,k)}{\Omega_N(M,2S)},
\end{equation}
where $\Omega_1^{+}(N,M,2S,k)$ is the number of all possible configurations with $N$ spins, magnetization $M$, $2S$ defects and $k$ spins $+1$ surrounded by a single defect, and $\Omega_N(M,2S)$ is the total number of configurations with $N$ spins, magnetization $M$ and defects $2S$. The expression of $\Omega_N(M,2S)$ has been derived in Ref.~\cite{antal2004probability} in terms of binomial coefficients, and it is given by
\begin{equation}\label{SMeq:bincoeff}
    \Omega_N(M,2S) = \binom{\frac{N+M}{2}-1}{S-1} \binom{\frac{N-M}{2}}{S} + \binom{\frac{N-M}{2}-1}{S-1} \binom{\frac{N+M}{2}}{S} + \delta_{M,\pm N}\delta_{S,0}   
\end{equation}
with $\delta$ here denoting the Kronecker delta. We have obtained an expression also for $\Omega_1^{+}(N,M,2S,k)$:
\begin{equation}
    \Omega^+_1(N,M,2S,2k) = \left\{\begin{array}{ll}
    \frac{2N}{N-M}\binom{\frac{N-M}{2}}{\frac{N+M}{2}} \quad \text{ if } \quad S=\frac{N+M}{2} \text{ and } k=0,\\[2mm]
    \frac{2N}{N-M} \binom{\frac{N-M}{2}-k}{S-k} \binom{\frac{N+M}{2}-S-1}{k-1}\binom{\frac{N-M}{2}}{k} \quad \text{ otherwise},
    \end{array}\right.
    \label{SMeq:conj1}
\end{equation}
and similar expressions for the other transition rates; see also Ref.~\cite{our_paper_long}.

\textit{Dynamics: Fokker-Planck and Langevin equations.}---In the large $N$ limit, we can expand the master equation (\ref{eq:dpdt}) in powers of $\epsilon=1/N$ (see Section VIII of \cite{van1992stochastic}). At first order in $\epsilon$, we obtain an associated Fokker-Planck equation that describes the time evolution of the rescaled probability density $p(m,s;t) = \lim_{N\rightarrow + \infty} \frac{N^2}{4} P_N(M = mN,2S = sN; tN)$. Notice that the factor $N^2/4$ arises because $M$ and $2S$ increase in steps of $2$. Here, the time has been rescaled and is expressed in units of Monte Carlo sweeps. After some computations~\cite{our_paper_long}, we obtain the Fokker-Planck equation:
\begin{eqnarray}\label{eq:Fokker-Planck}
    &&\frac{\partial p(m,s;t)}{\partial t} = \frac{\partial }{\partial m}\big[D_1(m,s) p(m,s;t)\big]
    + \frac{\partial }{\partial s}[D_2(m,s) p(m,s;t)\big]\nonumber\\
    && \qquad +\epsilon \frac{\partial^2}{\partial m^2} \Big[ D_{11}(m,s) p(m,s;t)\Big]
    +\epsilon \frac{\partial^2}{\partial s^2} \Big[ D_{22}(m,s) p(m,s;t)\Big]\nonumber\\
    && \qquad +2\epsilon \frac{\partial^2}{  \partial m \partial s}\Big[D_{12}(m,s) p(m,s;t)\Big],
\end{eqnarray}
where the drift coefficients are
\begin{eqnarray}
    D_1(m,s) &=& m+c_1 \tanh(x) +  c_2 \tanh(x_{-}) + c_3 \tanh(x_{+})\nonumber\\
    &+&K\epsilon\left( \frac{c_4}{\cosh^2(x)} +  \frac{c_5}{\cosh^2(x_{-})} + \frac{c_6}{\cosh^2(x_{+})}\right),\nonumber\\
    D_2(m,s) &=& 2s -1 + c_2 \tanh(x_{-})-c_3 \tanh(x_{+})\nonumber\\
    &+& K \epsilon \left( \frac{c_5}{\cosh^2(x_{-})} - \frac{c_6}{\cosh^2(x_{+})} \right),
\end{eqnarray}
while the diffusion coefficients are defined as
\begin{eqnarray}
    && D_{11}(m,s) = 1 - c_4 \tanh(x)
    -c_5 \tanh(x_{-}) - c_6 \tanh(x_{+}),\nonumber\\
    && D_{22}(m,s) = 1 + c_1 -c_5 \tanh(x_{-}) - c_6 \tanh(x_{+}),\nonumber\\
    && D_{12}(m,s) = -m -c_5 \tanh(x_{-}) + c_6\tanh(x_{+})
\end{eqnarray}
with $x := h+Km$ and $x_{\pm} := x \pm 2J$ ($T=1$). All the functions $c_i = c_i(m,s)$ with $i=1,\ldots,6$, reported in \cite{our_paper_long}, depend on $m$ and $s$. The correctness of our results can be also tested by taking the limit $J=0$, which corresponds to the Curie-Weiss model. In such a case, indeed, we recover the expression derived in Ref.~\cite{mori2010asymptotic}. Moreover, setting $\epsilon=0$, the stationary solution of \eqref{eq:Fokker-Planck} is $p(m,s)=\delta(m-m^*)\delta(s-s^*)$, where $m^*$ and $s^*$ are the equilibrium values of the macroscopic variables as given in Eq.~(\ref{eq:ms}). The effect of having a finite value of $\epsilon$ is the broadening of the Dirac $\delta$ solution.

From the Fokker-Planck equation (\ref{eq:Fokker-Planck}), one can obtain the associated Langevin equation for the stochastic vector $\bm{X}={(m,s)}^{\rm tr}$ (`{\rm tr}' denotes transposition)~\cite{our_paper_long}:
\begin{equation}\label{eq:Langevin_eq}
    d\bm{X}_t = \bm{D}(\bm{X}_t)dt + \sqrt{2\epsilon}\, \bm{B}(\bm{X}_t)\,d\bm{W}_{t} \,,
\end{equation}
where $\bm{D}(\bm{X}) = {-(D_1(m,s),D_2(m,s))}^{\rm tr}$ is the drift, $\bm{W}_t$ is a standard Brownian motion and $\bm{B}$ is the $2\times 2$ diffusion matrix
\begin{equation}
\begin{pmatrix}
    \sqrt{D_{11}} & 0\\
    \frac{D_{12}}{\sqrt{D_{11}}} & \sqrt{D_{22}-\frac{D_{12}^2}{D_{11}}}
\end{pmatrix}.
\end{equation}

\textit{Universal slowing down at both the critical line and the tricritical point.}---Dynamical scaling theory~\cite{FisherPRL1972,hohenberg1977theory,Barber1983,Privman1990Editor,wang1995study,ma2018modern} allows us to formulate the following finite-size scaling relation for the average absolute magnetization $\langle \vert m(t,N) \vert \rangle$:
\begin{equation}\label{eq:dynamical_scaling}
    \big\langle \left\vert m(t,N) \right\vert \big\rangle \sim 
    N^{-\Delta_m} g\left( t N^{-\frac{ \Delta_m }{ \lambda_m }} \right),
\end{equation}
where $g$ is a model-dependent function~\cite{ozeki2007nonequilibrium}.
From (\ref{eq:dynamical_scaling}), we obtain the scaling formulas (\ref{eq:scaling_relations}), by imposing $t \sim N^{z/d_u}$~\cite{hohenberg1977theory}.

\begin{figure}[t!]
    \centering
    \includegraphics[width=0.85\linewidth]{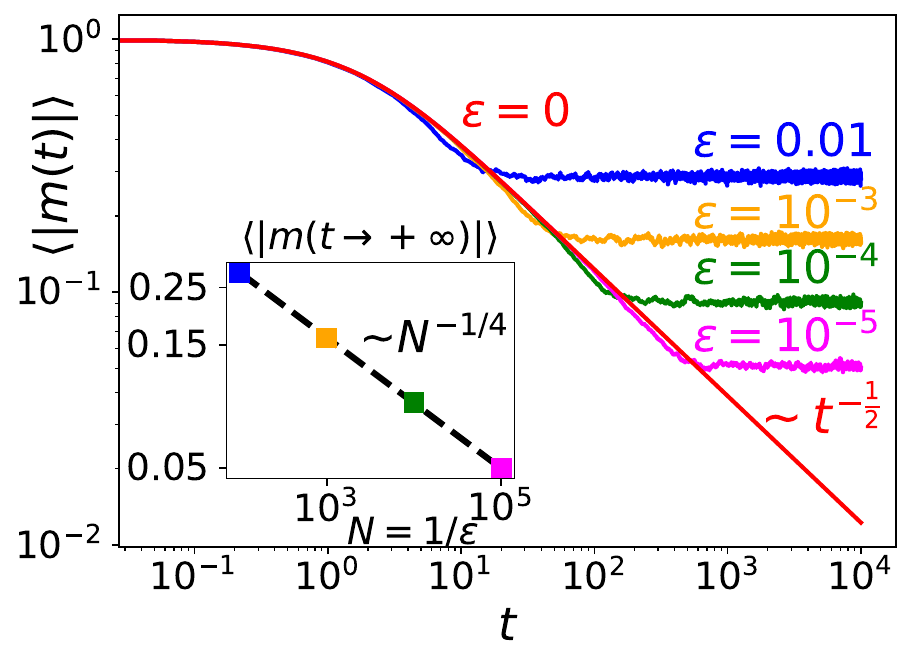}
    \includegraphics[width=0.85\linewidth]{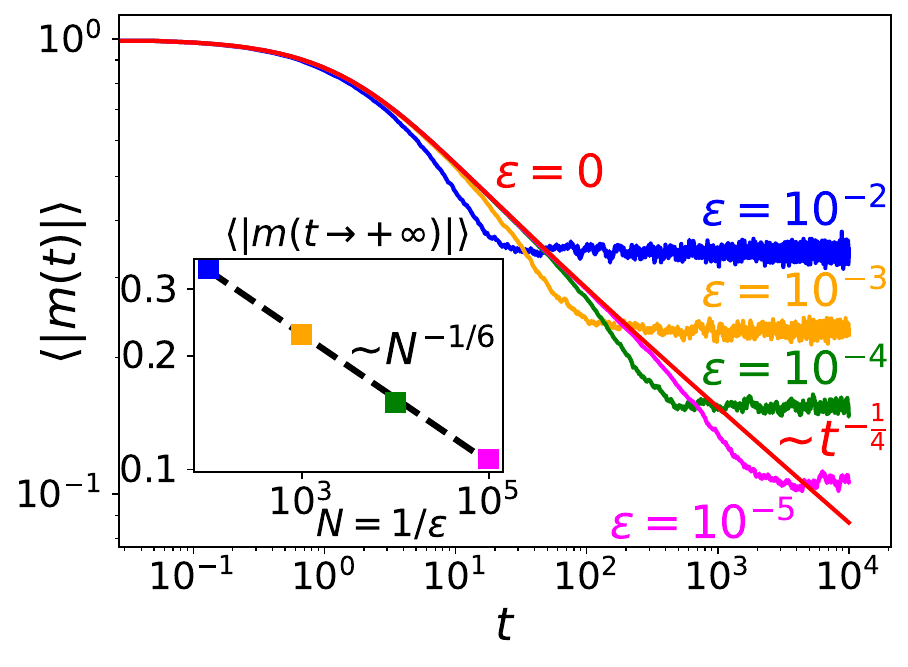}
    \caption{
    (Top) Time evolution of $\langle \vert m(t,N) \vert \rangle$, in log-log scale, at the critical point of the Curie-Weiss model, $J/T=0$, for decreasing values of $\epsilon=1/N$. We find that $z=2$ in an intermediate time range, before the magnetization reaches the equilibrium plateau proportional to $N^{-1/4}$ (see inset). (Bottom) Time evolution of $\langle \vert m(t,N) \vert \rangle$, in log-log scale, at the tricritical point of the Nagle-Kardar model [\eqref{eq:Hamiltonian}]. Also for this model $z=2$, and the asymptotic magnetization fluctuates around $N^{-1/6}$. In all simulations, the average is performed over $1000$ independent trajectories of the Langevin equation, \eqref{eq:Langevin_eq}, with $m(0)=1$.
    }
    \label{fig:powerlaw_scalings}
\end{figure}

On the critical line, it is known that the static critical exponents are those of the mean-field theory~\cite{Goldenfeld1992Lectures}, i.e., $\beta=1/2$ and $\nu=1/2$. To determine $\Delta_m$ using \eqref{eq:new_scaling_law}, one has to fix a value of $d_u$. Following~\cite{Ellis2010Asymptotic,ColonnaRomanoPRE2014}, we guess that the mean-field term in the Nagle-Kardar model dominates the critical  behavior, such that one can set $d_u=4$. Hence, from \eqref{eq:new_scaling_law}, we obtain $\Delta_m=1/4$. It coincides with the value that we compute by scaling the probability of magnetization with $N$, using large-deviation theory~\cite{our_paper_long}. The value $\Delta_m=1/4$ is confirmed numerically by looking at the scaling of $\langle \vert m(t,N) \vert \rangle$ with $N$ at large $t$ for the Curie-Weiss model; see the inset of Fig.~\ref{fig:powerlaw_scalings} in the top panel. We expect that this scaling holds at other points along the critical line.

At the tricritical point, we find that $\beta=1/4$~\cite{our_paper_long}. We know that the value of $\nu$ remains $1/2$ because it is related only to the coefficient that multiplies the kinetic term in the Landau free-energy~\cite{Goldenfeld1992Lectures}. Instead, the upper critical dimension changes to $d_u=3$~\cite{WegnerPRB1972}, as given by the Ginzburg criterion reported in Appendix A. Therefore, we obtian $\Delta_m=1/6$, as also confirmed numerically by looking at the long time behavior of $\langle \vert m(t,N) \vert \rangle$; see the inset of Fig.~\ref{fig:powerlaw_scalings}, Bottom panel. It is remarkable to observe that a similar change of the finite-size exponents from critical to tricritical also holds for the amplitude of the Casimir force~\cite{Dantchev2024}.

Let us now discuss the dynamical critical exponent $z$. First, we consider $J/T = 0$ and $K/T =1$, which corresponds to the critical point of the Curie-Weiss model~\cite{mori2010asymptotic}. In such a case, the time evolution of $m$ around criticality in the thermodynamic limit obeys the equation $\frac{dm}{dt} = m-\tanh(K m/T)$, whose integration at $K/T=1$ entails $m(t)\sim t^{-1/2}$ at large times. This gives $\lambda_m=1/2$, which implies $z=2$. We show this scaling in Fig.~\ref{fig:powerlaw_scalings} (Top panel), where we plot the time evolution of $\langle \vert m(t,N) \vert \rangle$ for the Curie-Weiss model, using the Langevin equation (\ref{eq:Langevin_eq}).

We have extended the analysis of $z$ to the Nagle-Kardar model. The asymptotic behavior of magnetization near criticality can be derived using a center manifold approach. While algorithms exist for constructing center manifolds for stochastic systems~\cite{roberts2014model}, here we can determine the exponent $z$ in the thermodynamic limit where dynamics becomes deterministic. In this limit, the center manifold construction simplifies~\cite{guckenheimer2013nonlinear} and yields the following nonlinear evolution equation for $m$ around criticality in the $m$-direction (see also Appendix B): 
\begin{eqnarray}\label{eq:m_dot}
    \frac{dm}{dt} &=& \frac{K^2 (K^2-3)}{3(1+K^2)} m^3 + O(m^5).
\end{eqnarray}
Using \eqref{eq:m_dot} we obtain $\lambda_m=1/2$ on the critical line, except at the tricritical point where the coefficient in front of $m^3$ vanishes. Since we have checked that the coefficient of $m^5$ is non-zero, $\lambda_m=1/4$ such that $z=2$. We have numerically verified this value in Fig.~\ref{fig:powerlaw_scalings} (Bottom panel). Notice that for all values of $K$ different from zero (even if very small), the critical exponents remain the same. Instead, according to the expansion in $m$ leading to \eqref{eq:m_dot}, the average absolute magnetization does not decay for $K=0$.

\textit{Conclusions.}---In this Letter, we have analyzed the Glauber dynamics of the Nagle-Kardar model. Our main goal was the derivation of the critical relaxation dynamics of the average absolute magnetization $\langle\vert m(t,N) \vert\rangle$, starting from a purely microscopic approach. We have been able to derive a Fokker-Planck equation for the magnetization  and the fraction of defects from the microscopic master equation. From the associated Langevin equation, we have obtained deterministic dynamical equations for the magnetization and the number of defects in the large $N$ limit. Then, using center manifold theory, we have extracted the dynamics of the magnetization, from which we have derived the value of the dynamical critical exponent $z=2$ that enters the scaling law $\left\langle \left\vert m(t) \right\vert \right\rangle \sim t^{-\lambda_m}$ of \eqref{eq:scaling_relations}. To reach such conclusion, we have shown that $\lambda_m=1/2$ with $\Delta_m=1/4$ on the critical line, while $\lambda_m=1/4$ with $\Delta_m=1/6$ at the tricritical point. As the dynamical critical exponent is $z=2$, we can affirm that the Nagle-Kardar model belongs to the dynamical universality class of purely relaxational non-conserved order parameter (model A).  

These findings may have profound implications for multiple areas of research. By incorporating nearest-neighbor interaction effects beyond the mean-field level, the Nagle-Kardar model allows for the study of competing interactions in spin-glass systems~\cite{mezard1987spin}. The model can also describe systems such as atoms in cavity~\cite{blass2018quantum}, magnetic chains~\cite{gatteschi2014single}, and fully coupled oscillators~\cite{gupta2017spontaneous}. In an interdisciplinary perspective, the Nagle-Kardar model has also been used to address some limitations of the Huxley-Simmons model of isometric muscle contraction~\cite{daRocha2021}. In this respect, its impact could also be extended to improve our knowledge about how nearest-neighbor interactions can enter the modeling of cooperativity phenomena in several fields of chemistry and biology, where only the Curie-Weiss model has been used (e.g., chemical kinetics~\cite{di2012mean}).

\textit{Appendix A: Ginzburg criterion for the upper critical dimension $d_u$}---The Nagle-Kardar lattice model can be mapped in the thermodynamic limit onto a continuum field theory in terms of a scalar field $\phi$. In units where the action is dimensionless, the scaling dimension of the field $\phi$ in space-dimension $d$ is $\Delta_\phi=(d-2)/2$. On the second-order critical line, the Landau potential is of order $\phi^4$, while at the tricritical point it is of order $\phi^6$. To compute $d_u$, we use the Ginzburg criterion, by equating the dimension of the Landau potential to $d$. Therefore, at the critical point, we have to solve the algebraic equation $4(d-2)/2=d$, which has the unique solution $d=d_u=4$. At the tricritical point, instead, the algebraic equation is $6(d-2)/2=d$ that is solved by $d=d_u=3$.

\textit{Appendix B: Derivation of \eqref{eq:m_dot} using center manifold theory}---Along the second-order phase transition line $K/T = e^{-2 J/T}$, the fraction of defects is $s^{*}=K/(K+1)$ and the magnetization is $m^{*}=0$. Close to this equilibrium, the dynamics of the deviations $\delta m \equiv x$ and $\delta s\equiv y$, up to terms of order $4$ in $x^py^q$, are given by
\begin{eqnarray}
    \frac{dx}{dt} = \dot{x} &=& (ay+by^2)x + (c+dy+ey^2)x^3, 
    \label{eq:dx_dt}\\
    \frac{dy}{dt} = \dot{y} &=& Ay + By^2 + (C+Dy+Ey^2)x^2 + F x^{4},\label{eq:dy_dt}
\end{eqnarray}
where the coefficients $a,b,\ldots,e$, $A,B,\ldots,F$ depend only on the strength $K$ of the mean-field coupling. In \eqref{eq:dx_dt},
\begin{eqnarray*}
    a &=& -\frac{2 K(1+K){(K-1)}^3}{{(1+K^2)}^2}, \quad 
    b = -2K + \frac{8K^3}{{(1+K^2)}^2},\\
    c &=& -\frac{2 K^3(3+K-3K^2+K^3)}{3{(1+K^2)}^2},
\end{eqnarray*}
while in \eqref{eq:dy_dt},
\begin{eqnarray*}    
    A &=& -\frac{4K}{1+K^2}, \quad B = 2-\frac{4}{1+K^2}, \quad
    C = -2 + \frac{2}{1+K^2},\\
    D &=& -\frac{4\left[K+K^2(-1+2K-3K^3+K^4)\right]}{{(1+K^2)}^3}.
\end{eqnarray*}
We do not provide the expression of the coefficients $e,E,F$ as they do not enter the derivations below.

The system of equations (\ref{eq:dx_dt})-(\ref{eq:dy_dt}) can be rewritten as
\begin{equation}\label{eq:EM-system_equations}
    \begin{pmatrix}
\dot{x} \\[6pt]
\dot{y}
\end{pmatrix}
=
\begin{pmatrix}
0 & 0 \\[6pt]
0 & A
\end{pmatrix}
\begin{pmatrix}
x \\[6pt]
y
\end{pmatrix}
+
\begin{pmatrix}
(ay+by^2)x + (c+dy+ey^2)x^3 + fx^5\\[6pt]
By^2 + (C+Dy+Ey^2)x^2 + (F+Gy)x^4 
\end{pmatrix}.
\end{equation}
The eigenvalues of the linearized system are $\lambda_1=0$ and $\lambda_2=A<0$. This means that the equilibrium point $(0,0)$ is non-hyperbolic and we need to invoke the center manifold theory~\cite{guckenheimer2013nonlinear} to make further progress. The center manifold $E^c$ of the linearized system is the $x$-axis, as follows by computing the eigenvector corresponding to the eigenvalue $\lambda_1=0$. Hence, we parametrize the center manifold as $W^c=\Big\{(x,y):y=h(x),h(0)=0, h'(0)=0\Big\}$. The condition $h(0)=0$ comes from the fact that the center manifold passes through the origin, while the condition $h'(0)=0$ is due to the fact that $W^c$ is tangential to $E^c$ at the origin $(0,0)$. To determine $h(x)$, we express it in Taylor expansion by setting $y=h(x)=\alpha x^2 + \gamma x^3 + \chi x^4 + \zeta x^5 + O(x^6)$. On the one hand, from \eqref{eq:dy_dt}, we get
\begin{eqnarray}\label{y_dot_1}
    \dot{y} &=& (A\alpha + C) x^2 + A \gamma x^3 + (F+D\alpha + B\alpha^2 + A \chi)x^4 + \nonumber\\
    &+& (D\gamma + 2B\alpha\gamma + A \zeta)x^5 + O(x^6).
\end{eqnarray}
On the other hand, from \eqref{eq:dx_dt}, we obtain
\begin{equation}\label{y_dot_2}
    \dot{y} = h'(x)\dot{x}=(2c\alpha+2a \alpha^2)x^4 + (3c\gamma + 5a \alpha\gamma )x^{5}.
\end{equation}
Therefore, equating Eqs.~(\ref{y_dot_1})-(\ref{y_dot_2}), we deduce that 
\begin{eqnarray*}
    &&A\gamma = 0 \quad \Rightarrow \quad \gamma=0 ~(A \neq 0),\\
    &&A\alpha+C =0 \quad \Rightarrow \quad \alpha=-C/A,\\
    &&F+D\alpha + B\alpha^2 + A\chi = 2c\alpha + 2a \alpha^{2},\\
    &&D\gamma + 2B\alpha\gamma + A \zeta = 3c\gamma + 5a\alpha \gamma.
\end{eqnarray*}
At order $3$ in $x$, we get $y=h(x)=-\frac{C}{A}x^2 + O(x^4) = -\frac{K}{2}x^2 + O(x^4)$. Therefore, it finally follows that
\begin{equation}
    \dot{x} = \Big(c-a\frac{C}{A}\Big)x^3 + O(x^5) = \frac{K^2 (K^2-3)}{3(1+K^2)} x^3 + O(x^5).
\end{equation}

\begin{acknowledgments}
\textit{Acknowledgments.}---The authors would like to thank David Mukamel for carefully reading the manuscript and providing us with helpful comments. We acknowledge financial support from the PRIN 2022 PNRR P2022XPT32 ``Regulation of striated muscle: a research bridging single molecule to organ (ReStriMus)'' funded by the European Union---Next Generation EU, and from the MUR PRIN2022 project BECQuMB Grant No.~20222BHC9Z.
\end{acknowledgments}

\bibliography{bibliography.bib}

\end{document}